\documentclass[11pt]{article}
\usepackage[margin=1in]{geometry}
\usepackage[numbers,sort&compress]{natbib}
\usepackage{amsmath,amssymb}
\usepackage{graphicx}
\usepackage{tikz-cd}
\usepackage{multicol}

\title{Hawking Radiation and Greybody Factors of Test Scalar and Electromagnetic Fields on Asymptotically Flat Pure Lovelock Black Holes}
\author{Jayden Tan\\Equilibrium Research Unit, Analysis of Systemic Complexity\\\#02-01, 99 Portsdown Road, Singapore 139299}
\date{jaydtan2002@proton.me}

\begin{document}
\maketitle

\begin{abstract}
Pure Lovelock black holes are geometrically more transparent than their Einstein counterparts, but they radiate far less. We compute scalar and higher-dimensional electromagnetic greybody factors and Hawking spectra on the critical branch $d=2N+2$, compare them with Schwarzschild--Tangherlini black holes at the same horizon radius $r_h$, and show that the smaller Hawking temperature overwhelms the enhanced transmission. In the benchmark $d=6$ case, the integrated scalar and electromagnetic powers are reduced by about $10^{-3}$ and $10^{-5}$, respectively. We also find a clean higher-curvature signature: as the Lovelock order $N$ grows, Hawking radiation becomes increasingly dominated by the scalar-type electromagnetic sector.
\end{abstract}

\section{Introduction}

Black holes are not perfect black bodies. Although quantum fields on a stationary black-hole background lead to Hawking radiation with a thermal spectrum at the horizon, the flux measured by an observer at infinity is distorted by curvature-induced scattering outside the horizon. The frequency-dependent transmission coefficients through this potential barrier are the greybody factors, and they govern the absorption probability, the partial-wave emission rates and, in the low-energy regime, the absorption cross section of the geometry \cite{Hawking1975,Page1976,Unruh1976,DasGibbonsMathur1997}. Because they are simultaneously sensitive to near-horizon thermodynamics and to the global structure of the effective potential, greybody factors provide a clean probe of how departures from Einstein gravity modify black-hole evaporation.

Among higher-curvature theories, Lovelock gravity occupies a special place. It is the unique metric theory built from higher powers of the curvature while still yielding second-order field equations \cite{Lovelock1971}. A particularly economical sector is \emph{pure} Lovelock gravity, in which the action contains only a single Lovelock density of order $N$, with $N$ labeling the power of the curvature tensor kept in the action, with no Einstein--Hilbert term. Throughout the paper, $d$ denotes the spacetime dimension. This theory is singled out by several structural properties: the vacuum is kinematic in the odd critical dimension $d=2N+1$, while non-trivial vacuum dynamics begins in $d=2N+2$ \cite{DadhichGhoshJhingan2012}; moreover, bound orbits survive in the even critical dimensions $d=2N+2$, making pure Lovelock black holes the closest higher-dimensional analogues of the Schwarzschild geometry \cite{DadhichGhoshJhingan2013}. Their perturbative stability has also been investigated for the asymptotically flat branch \cite{GannoujiRodriguezDadhich2019}.

Pure Lovelock theory is especially interesting because it provides a clean way to isolate genuinely higher-curvature gravitational physics without mixing it with the Einstein term. In that sense it is not merely a technical truncation of Lovelock gravity, but a sharply defined alternative dynamical framework whose critical dimensions retain several features that make black-hole physics intuitive and physically interpretable: a nontrivial vacuum, bound orbits, and a simple one-scale black-hole geometry. This makes pure Lovelock an unusually good laboratory for asking which aspects of scattering and Hawking radiation are robust consequences of horizon geometry and which are specific to Einstein gravity.

For the problem at hand, the relevant background is the static, spherically symmetric, asymptotically flat vacuum solution of order-$N$ pure Lovelock gravity. In the conventions of Cai and Ohta, and equivalently in the form emphasized in later pure Lovelock work, the line element can be written as
\begin{equation}
\label{eq:pl_metric}
ds^2=-f(r)\,dt^2+\frac{dr^2}{f(r)}+r^2 d\Omega_{d-2}^2,
\end{equation}
where $t$ is the static time measured by an observer at infinity, $r$ is the areal radius, and $d\Omega_{d-2}^2$ is the line element on the unit $(d-2)$-sphere.
with
\begin{equation}
\label{eq:pl_lapse}
f(r)=1-\left(\frac{\mu}{r^{d-2N-1}}\right)^{1/N},
\qquad d\geq 2N+2.
\end{equation}
The function $f(r)$ is the lapse, and $\mu$ is an integration constant proportional to the black-hole mass. In alternative conventions one often writes the mass parameter as $M$; throughout this paper, $r_h$ denotes the event-horizon radius and therefore the natural length scale of the problem. Different papers absorb numerical factors and the Lovelock coupling into the mass parameter; one therefore also encounters the equivalent parametrizations $f(r)=1-(2M/r^{d-2N-1})^{1/N}$ or $f(r)=1-(r_h/r)^{(d-2N-1)/N}$ \cite{CaiOhta2006,DadhichPonsPrabhu2013}. The horizon radius is fixed by $f(r_h)=0$, so that $\mu=r_h^{\,d-2N-1}$. In the even critical dimension $d=2N+2$, which is the most distinctive pure Lovelock setting, Eq.~\eqref{eq:pl_lapse} reduces to
\begin{equation}
\label{eq:pl_lapse_critical}
f(r)=1-\left(\frac{\mu}{r}\right)^{1/N}=1-\left(\frac{r_h}{r}\right)^{1/N}.
\end{equation}
This is the asymptotically flat Schwarzschild analogue of pure Lovelock gravity. It isolates higher-curvature effects without contamination from the Einstein term.

On the Einstein side, greybody factors and Hawking spectra are well studied for higher-dimensional Schwarzschild--Tangherlini black holes, including exact scalar-field calculations and bulk photon emission in which the Maxwell field splits into scalar-type and vector-type sectors on $S^{d-2}$ \cite{HarrisKanti2003,JungPark2007}. By contrast, the higher-curvature literature is concentrated mainly on generic Lovelock or Einstein--Gauss--Bonnet backgrounds rather than asymptotically flat pure Lovelock vacuum black holes \cite{ChenLiaoWang2011,ZhangLiChen2018,LiZhang2019}. The scalar sector has received some attention in broader Lovelock settings, but a dedicated study of test scalar and higher-dimensional electromagnetic greybody factors on asymptotically flat pure Lovelock black holes still appears to be absent.

At the same time, black-hole perturbations have become one of the main diagnostic tools in higher-curvature gravity. In four-dimensional higher-derivative settings, quasinormal spectra, analytic metric reconstructions and Hawking fluxes have already been used to characterize non-Schwarzschild solutions and their observational signatures \cite{Konoplya:2022iyn,Konoplya:2019ppy,Kokkotas:2017zwt}. Similar ideas have been extended to geometries supported by infinite towers of curvature corrections and to Einstein-dilaton/scalar--Gauss--Bonnet black holes, where accurate metric approximations enable calculations of shadows and related observables \cite{Konoplya:2024kih,Konoplya:2019fpy,Kokkotas:2017ymc}. In Gauss--Bonnet and Lovelock gravity itself, the perturbation literature already covers scalar, tensor, vector and fermionic sectors, together with asymptotic, large-$D$ and dimension-dependent quasinormal spectra \cite{Abdalla:2005hu,Prasobh:2014zea,Chakrabarti:2005cm,Chakrabarti:2006ei,Chen:2015fuf,Yoshida:2015vua}. Additional studies explored asymptotic-mode behavior, clouds of strings, fermionic and tensor perturbations, and other nontrivial matter or compactification effects \cite{Gonzalez:2018xrq,Churilova:2019sah,Cao:2023mai,Daghigh:2006xg,Sadeghi:2011zza,MoraisGraca:2016hmv}.

Recent greybody-factor work has also grown rapidly around the quasinormal-mode/greybody-factor correspondence: Malik studied Euler--Heisenberg electrodynamics, Morris--Thorne wormholes, supermassive black holes with dark matter halos and Bonanno--Reuter regular black holes \cite{Malik2025EEH,Malik2024MorrisThorne,Malik2025DMHalo,Malik2025BRMassive}; Lütfüoğlu studied Proca--Gauss--Bonnet, Weyl-gravity, non-minimal Einstein--Yang--Mills, proper-time asymptotic-safety, Dehnen-halo and asymptotically safe backgrounds \cite{Lutfuoglu2025PGB,Lutfuoglu2025Weyl,Lutfuoglu2025EYM,LutfuogluEtAl2025PT,Lutfuoglu2025Dehnen,Lutfuoglu2026ASG}; Bolokhov and Skvortsova considered Bonanno--Reuter and Hayward regular black holes, traversable wormholes, Einasto-supported regular black holes and quantum-corrected geometries \cite{BolokhovSkvortsova2025BR,BolokhovSkvortsova2024WH,Bolokhov2026Einasto,BolokhovSkvortsova2026Hayward,Skvortsova2025QuantumCorr}; and Dubinsky analyzed quasi-topological, GMGHS, parametrized and quantum-corrected Hayward black holes \cite{Dubinsky2026QTG,Dubinsky2025GMGHS,DubinskyZinhailo2025AnalyticGBF,Dubinsky2026Hayward}. The basic correspondence between quasinormal modes and grey-body factors was formulated by Konoplya and Zhidenko for spherically symmetric black holes and later extended to rotating geometries \cite{KonoplyaZhidenko2024Corr,KonoplyaZhidenko2025RotCorr}.

A second lesson from this broader literature is that higher-curvature black holes can display new instabilities, nonperturbative branches and finite-coupling effects, both in de Sitter/AdS backgrounds and in more general Lovelock settings \cite{Cuyubamba:2016cug,Konoplya:2017ymp,Konoplya:2017zwo,Konoplya:2017lhs,Grozdanov:2016fkt,Mishra:2020gce}. Related work has also addressed string-corrected stability and absorption, long-lived or wormhole configurations, lower-dimensional Einstein--Lovelock solutions, and effective higher-order curvature corrections beyond Gauss--Bonnet \cite{Cuyubamba:2018jdl,Konoplya:2010vz,Konoplya:2020ibi,Konoplya:2020der,Moura:2006pz,Matyjasek:2020bzc}. In Einstein-dilaton/scalar--Gauss--Bonnet models, axial, polar, radial and rapidly rotating perturbations have been analyzed in detail \cite{Pierini:2022eim,Blazquez-Salcedo:2024oek,Blazquez-Salcedo:2022omw,Staykov:2021dcj,Blazquez-Salcedo:2020caw,Blazquez-Salcedo:2020rhf}. These developments now connect directly with observation, where inspiral and ringdown measurements are used to probe higher-curvature corrections and black-hole spectroscopy in concrete modified-gravity models \cite{Carson:2020ter,Cano:2020cao,Wagle:2021tam,Silva:2022srr}.

In this paper we therefore extend the greybody analysis of asymptotically flat pure Lovelock black holes to both a minimally coupled test scalar field and a Maxwell field propagating in the full spacetime dimension. Our goals are threefold. First, we determine how the transmission coefficients depend on the Lovelock order $N$, the spacetime dimension $d$, the angular momentum number $\ell$ and the frequency $\omega$. Second, we compare the pure Lovelock results with those of Schwarzschild--Tangherlini black holes in the same dimension in order to separate genuinely pure Lovelock effects from generic higher-dimensional scattering effects. Third, we identify the distinctive electromagnetic signatures that arise because the higher-dimensional photon field contains separate scalar-type and vector-type channels. As we will see, pure Lovelock black holes are simultaneously more transparent and much colder than their Tangherlini counterparts, and this competition leaves a characteristic imprint on the Hawking spectra.

The paper is organized so that the physical story unfolds continuously from geometry to radiation. Section~2 fixes the field equations, notation and mode degeneracies; Section~3 presents the numerical scattering data and frequency-resolved spectra; Section~4 converts those spectra into integrated power estimates; and Section~5 summarizes the common picture that emerges from all three levels of description.

\section{Test-Field Equations and Scattering Setup}

Throughout this section, ``test field'' means that the scalar and electromagnetic perturbations propagate on the fixed black-hole background and do not backreact on the metric.

A minimally coupled test scalar field on the background \eqref{eq:pl_metric} satisfies
\begin{equation}
\Box\Phi=0.
\end{equation}
Here $\Box$ is the covariant d'Alembertian associated with the metric \eqref{eq:pl_metric}.
After the standard separation
\begin{equation}
\Phi(t,r,\Omega)=e^{-i\omega t}Y_{\ell}(\Omega)\,r^{-(d-2)/2}\,\psi_{\ell}(r),
\end{equation}
where $\Omega$ collectively denotes the angular coordinates on $S^{d-2}$, $Y_{\ell}$ is the scalar spherical harmonic, $\ell$ labels the angular-momentum multipole, $\omega$ is the mode frequency measured at infinity, and $\psi_{\ell}(r)$ is the radial mode function. With the introduction of the tortoise coordinate $dr_\ast=dr/f(r)$, which stretches the exterior region so that the horizon is pushed to $r_\ast\to-\infty$, the radial equation takes the one-dimensional Schr\"odinger form
\begin{equation}
\label{eq:scalar_schrodinger}
\frac{d^2\psi_{\ell}}{dr_\ast^2}+\bigl[\omega^2-V^{(0)}_{\ell}(r)\bigr]\psi_{\ell}=0,
\end{equation}
with effective potential
\begin{equation}
\label{eq:scalar_potential_general}
V^{(0)}_{\ell}(r)=f(r)\left[\frac{\ell(\ell+d-3)}{r^2}+\frac{(d-2)f'(r)}{2r}+\frac{(d-2)(d-4)}{4r^2}f(r)\right].
\end{equation}

For a test Maxwell field $A_\mu$, the field equations are
\begin{equation}
\nabla_\mu F^{\mu\nu}=0,
\qquad F_{\mu\nu}=\partial_\mu A_\nu-\partial_\nu A_\mu.
\end{equation}
Here $A_\mu$ is the vector potential, $F_{\mu\nu}$ its field-strength tensor, and $\nabla_\mu$ the metric-compatible covariant derivative. Following the Kodama--Ishibashi decomposition on the sphere $S^{d-2}$, the physical electromagnetic perturbations separate into scalar-type and vector-type sectors \cite{KodamaIshibashi2004,JungPark2007}. The label $I=S,V$ distinguishes these two families of photon polarizations on the sphere. Each sector is governed by a master field $\Psi^{(I)}_{\ell}$ satisfying
\begin{equation}
\label{eq:em_schrodinger}
\frac{d^2\Psi^{(I)}_{\ell}}{dr_\ast^2}+\bigl[\omega^2-V^{(I)}_{\ell}(r)\bigr]\Psi^{(I)}_{\ell}=0,
\qquad I=S,V,
\end{equation}
with
\begin{equation}
\label{eq:em_scalar_potential}
V^{(S)}_{\ell}(r)=f(r)\left[\frac{\ell(\ell+d-3)}{r^2}+\frac{(d-2)(d-4)}{4r^2}f(r)-\frac{(d-4)f'(r)}{2r}\right],
\qquad \ell\geq 1,
\end{equation}
and
\begin{equation}
\label{eq:em_vector_potential}
V^{(V)}_{\ell}(r)=f(r)\left[\frac{(\ell+1)(\ell+d-4)}{r^2}+\frac{(d-4)(d-6)}{4r^2}f(r)+\frac{(d-4)f'(r)}{2r}\right],
\qquad \ell\geq 1.
\end{equation}
In four dimensions the two electromagnetic sectors coincide, but for $d>4$ they split: the scalar-type channel is systematically less suppressed than the vector-type channel because of the relative sign of the $f'(r)$ term.

For any of the one-dimensional barrier problems above, the physically relevant boundary conditions are imposed in the tortoise coordinate $r_\ast$, for which the horizon is sent to $r_\ast\to -\infty$ and spatial infinity to $r_\ast\to +\infty$. Choosing unit incoming flux from infinity, one writes
\begin{equation}
\label{eq:gbf_bc}
\Psi_\ell(r_\ast)\sim
\begin{cases}
\mathcal{T}_\ell(\omega)e^{-i\omega r_\ast}, & r_\ast\to -\infty,\\[1ex]
e^{-i\omega r_\ast}+\mathcal{R}_\ell(\omega)e^{i\omega r_\ast}, & r_\ast\to +\infty.
\end{cases}
\end{equation}
Here $\Psi_\ell$ denotes whichever radial master field is under discussion, and $\mathcal{T}_\ell$ and $\mathcal{R}_\ell$ are the transmission and reflection amplitudes, respectively. The absence of an outgoing wave at the horizon expresses the black-hole boundary condition, while the asymptotic form at infinity contains the incident and reflected components. The greybody factor is then defined as the transmission probability
\begin{equation}
\label{eq:gbf_definition}
\gamma_\ell(\omega)=\lvert\mathcal{T}_\ell(\omega)\rvert^2=1-\lvert\mathcal{R}_\ell(\omega)\rvert^2,
\end{equation}
where the second equality follows from flux conservation for a real single-barrier potential.

In this work the greybody factors are obtained by direct numerical integration of the master equations in the tortoise coordinate. For each field sector and each partial wave, we first construct the effective potential on a numerical radial grid, compute the tortoise map $r_\ast(r)$ from $dr_\ast=dr/f(r)$, and interpolate $V_\ell$ as a function of $r_\ast$. Since the horizon corresponds to $r_\ast\to -\infty$, the integration is started at a small offset from the horizon where the exact purely ingoing behavior is already well approximated by
\begin{equation}
\label{eq:numerical_horizon_bc}
\Psi_\ell(r_\ast)\propto e^{-i\omega r_\ast},
\qquad
\frac{d\Psi_\ell}{dr_\ast}=-i\omega\Psi_\ell.
\end{equation}
Because the radial equation is linear, the overall normalization is arbitrary, and we may therefore choose the initial data
\begin{equation}
\label{eq:numerical_initial_data}
\Psi_\ell(r_{\ast,\min})=1,
\qquad
\left.\frac{d\Psi_\ell}{dr_\ast}\right|_{r_{\ast,\min}}=-i\omega,
\end{equation}
at a sufficiently negative tortoise coordinate $r_{\ast,\min}$.

The numerical solution is then integrated outward to a large positive tortoise coordinate, where the potential has decayed and the wave takes the asymptotic free form
\begin{equation}
\label{eq:numerical_asymptotic_fit}
\Psi_\ell(r_\ast)\simeq A^{\mathrm{in}}_\ell(\omega)e^{-i\omega r_\ast}+A^{\mathrm{out}}_\ell(\omega)e^{i\omega r_\ast}.
\end{equation}
Combining Eq.~\eqref{eq:numerical_asymptotic_fit} with its derivative gives the amplitude-extraction formulas
\begin{equation}
\label{eq:numerical_amplitudes}
A^{\mathrm{in}}_\ell=\frac{1}{2}e^{i\omega r_\ast}\left(\Psi_\ell+\frac{i}{\omega}\frac{d\Psi_\ell}{dr_\ast}\right),
\qquad
A^{\mathrm{out}}_\ell=\frac{1}{2}e^{-i\omega r_\ast}\left(\Psi_\ell-\frac{i}{\omega}\frac{d\Psi_\ell}{dr_\ast}\right).
\end{equation}
In practice we evaluate these expressions over the far-zone tail of the numerical solution and average them there in order to reduce the residual oscillations produced by finite-radius extraction.

The normalization in Eq.~\eqref{eq:numerical_initial_data} fixes the ingoing horizon amplitude rather than the incident amplitude at infinity. To recover the convention of Eq.~\eqref{eq:gbf_bc}, one rescales the entire solution by $1/A^{\mathrm{in}}_\ell$. The transmission and reflection amplitudes are therefore
\begin{equation}
\label{eq:numerical_TR}
\mathcal{T}_\ell(\omega)=\frac{1}{A^{\mathrm{in}}_\ell(\omega)},
\qquad
\mathcal{R}_\ell(\omega)=\frac{A^{\mathrm{out}}_\ell(\omega)}{A^{\mathrm{in}}_\ell(\omega)},
\end{equation}
and the greybody factor follows as
\begin{equation}
\label{eq:numerical_gamma}
\gamma_\ell(\omega)=\frac{1}{\lvert A^{\mathrm{in}}_\ell(\omega)\rvert^2},
\qquad
\lvert\mathcal{R}_\ell(\omega)\rvert^2=\left\lvert\frac{A^{\mathrm{out}}_\ell(\omega)}{A^{\mathrm{in}}_\ell(\omega)}\right\rvert^2.
\end{equation}
This direct integration and matching procedure is the one used throughout the paper to generate the greybody factors and, through them, the absorption and Hawking-emission observables discussed below.

For the scalar field, the absorption cross section is obtained by summing the partial-wave contributions,
\begin{equation}
\label{eq:total_sigma_abs}
\sigma_{\mathrm{abs}}(\omega)=\sum_{\ell=0}^{\infty}\sigma_\ell(\omega),
\end{equation}
with
\begin{equation}
\label{eq:partial_sigma_abs}
\sigma_\ell(\omega)=\frac{2^{d-4}\pi^{\frac{d-3}{2}}\Gamma\!\left(\frac{d-3}{2}\right)}{(d-4)!\,\omega^{d-2}}\,\frac{(2\ell+d-3)(\ell+d-4)!}{\ell!}\,\gamma_\ell(\omega).
\end{equation}
In the plots below it is convenient to display the dimensionless reduced partial absorption cross sections
\begin{equation}
\label{eq:reduced_sigma_abs}
\hat{\sigma}_\ell(\omega)=\frac{(\omega r_h)^{d-2}}{r_h^{d-2}}\,\sigma_\ell(\omega),
\end{equation}
which remove the universal kinematic factor $\omega^{-(d-2)}$.

The same greybody factors enter the Hawking emission rates. For the scalar field,
\begin{equation}
\label{eq:scalar_emission}
\frac{d^2E_{\mathrm{sc}}}{dt\,d\omega}=\frac{1}{2\pi}\sum_{\ell=0}^{\infty}D_{\ell}^{(0)}\,\frac{\omega\,\gamma_\ell(\omega)}{e^{\omega/T_H}-1},
\end{equation}
where the scalar-harmonic multiplicity on $S^{d-2}$ is
\begin{equation}
\label{eq:scalar_deg}
D_{\ell}^{(0)}=\frac{(2\ell+d-3)(\ell+d-4)!}{\ell!\,(d-3)!}.
\end{equation}
For the electromagnetic field,
\begin{equation}
\label{eq:em_emission}
\frac{d^2E_{\mathrm{EM}}}{dt\,d\omega}=\frac{1}{2\pi}\sum_{\ell=1}^{\infty}\left[D_{\ell}^{(S)}\,\frac{\omega\,\gamma^{(S)}_\ell(\omega)}{e^{\omega/T_H}-1}+D_{\ell}^{(V)}\,\frac{\omega\,\gamma^{(V)}_\ell(\omega)}{e^{\omega/T_H}-1}\right],
\end{equation}
with
\begin{equation}
\label{eq:em_degeneracies}
D_{\ell}^{(S)}=D_{\ell}^{(0)},
\qquad
D_{\ell}^{(V)}=\frac{\ell(\ell+d-3)(2\ell+d-3)(\ell+d-5)!}{(\ell+1)!\,(d-4)!}.
\end{equation}
In Eqs.~\eqref{eq:scalar_emission} and \eqref{eq:em_emission}, $d^2E/(dt\,d\omega)$ is the energy emitted per unit asymptotic time $t$ and frequency $\omega$, $T_H=f'(r_h)/(4\pi)$ is the Hawking temperature set by the surface gravity, and the factors $D_{\ell}^{(0)}$, $D_{\ell}^{(S)}$ and $D_{\ell}^{(V)}$ count the multiplicity of scalar and vector harmonics on $S^{d-2}$. We will emphasize the emissivity rather than an electromagnetic absorption cross section because the two physical sectors enter with different harmonic multiplicities, and the Hawking spectrum exposes their competition most directly. This notation will be used uniformly below: we first analyze transmission, then frequency-resolved emission, and finally integrated power.

\section{Numerical Results}

We now turn to explicit numerical solutions. The discussion is ordered from the simplest to the richest problem: first the scalar benchmark, then the comparison along the critical pure Lovelock sequence, next the Tangherlini reference geometry, and finally the split electromagnetic sector. This ordering keeps the paper from reading like disconnected case studies and makes the common physical mechanism---enhanced transmission versus reduced Hawking temperature---easy to track throughout.

To display the pure Lovelock dependence as transparently as possible, we focus first on the even critical dimensions $d=2N+2$. On this branch the metric function simplifies to
\begin{equation}
\label{eq:critical_lapse}
f_N(r)=1-\left(\frac{r_h}{r}\right)^{1/N},
\qquad d=2N+2,
\end{equation}
where the subscript $N$ reminds us that the geometry is fixed by the Lovelock order.
and the scalar-field potential can be written as
\begin{equation}
\label{eq:critical_potential}
V^{(N)}_{\ell}(r)=\frac{f_N(r)}{r^2}\left[\ell(\ell+2N-1)+N(N-1)+\bigl(1-N(N-1)\bigr)\left(\frac{r_h}{r}\right)^{1/N}\right].
\end{equation}
In all numerical examples we set $r_h=1$, so every frequency is quoted through the dimensionless combination $\omega r_h$. For each field sector and each partial wave, we integrate the relevant Schr\"odinger equation outward from near-horizon ingoing data and extract the asymptotic incoming and outgoing amplitudes from the far-zone tail as described in Section~2. The resulting $\gamma_\ell(\omega)$ is then used directly in the greybody-factor, absorption and Hawking-emission plots below. The numerical consistency check $\gamma_{\ell}+|\mathcal{R}_{\ell}|^2=1$ is satisfied to better than $1.6\times 10^{-4}$ for all curves shown below.

\subsection{Benchmark Scalar Case: $N=2$ in $d=6$}

\begin{figure}[t]
\centering
\includegraphics[width=0.82\textwidth]{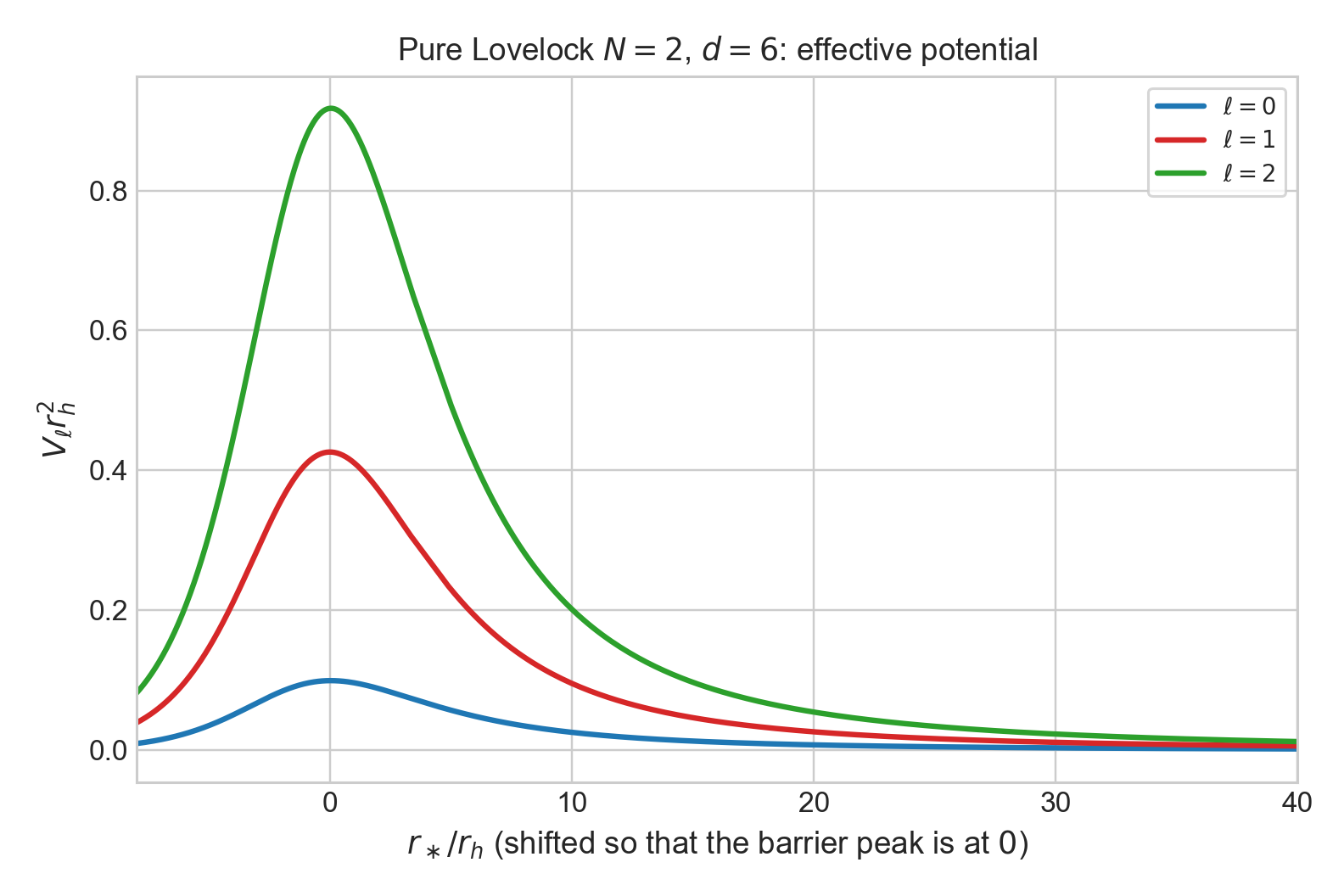}
\caption{Effective scalar potentials for the pure Lovelock black hole with $N=2$ and $d=6$, shown for $\ell=0,1,2$. The tortoise coordinate is shifted so that the barrier peak of each mode is located at the origin.}
\label{fig:n2_d6_potential}
\end{figure}

We begin with the lowest nontrivial pure Lovelock order, $N=2$, for which the critical dimension is $d=6$. Figure~\ref{fig:n2_d6_potential} shows the effective potential for the first three scalar partial waves. As expected, the barrier becomes both higher and broader as $\ell$ increases, so low-frequency transmission is progressively suppressed for higher multipoles.

\begin{figure}[t]
\centering
\includegraphics[width=0.82\textwidth]{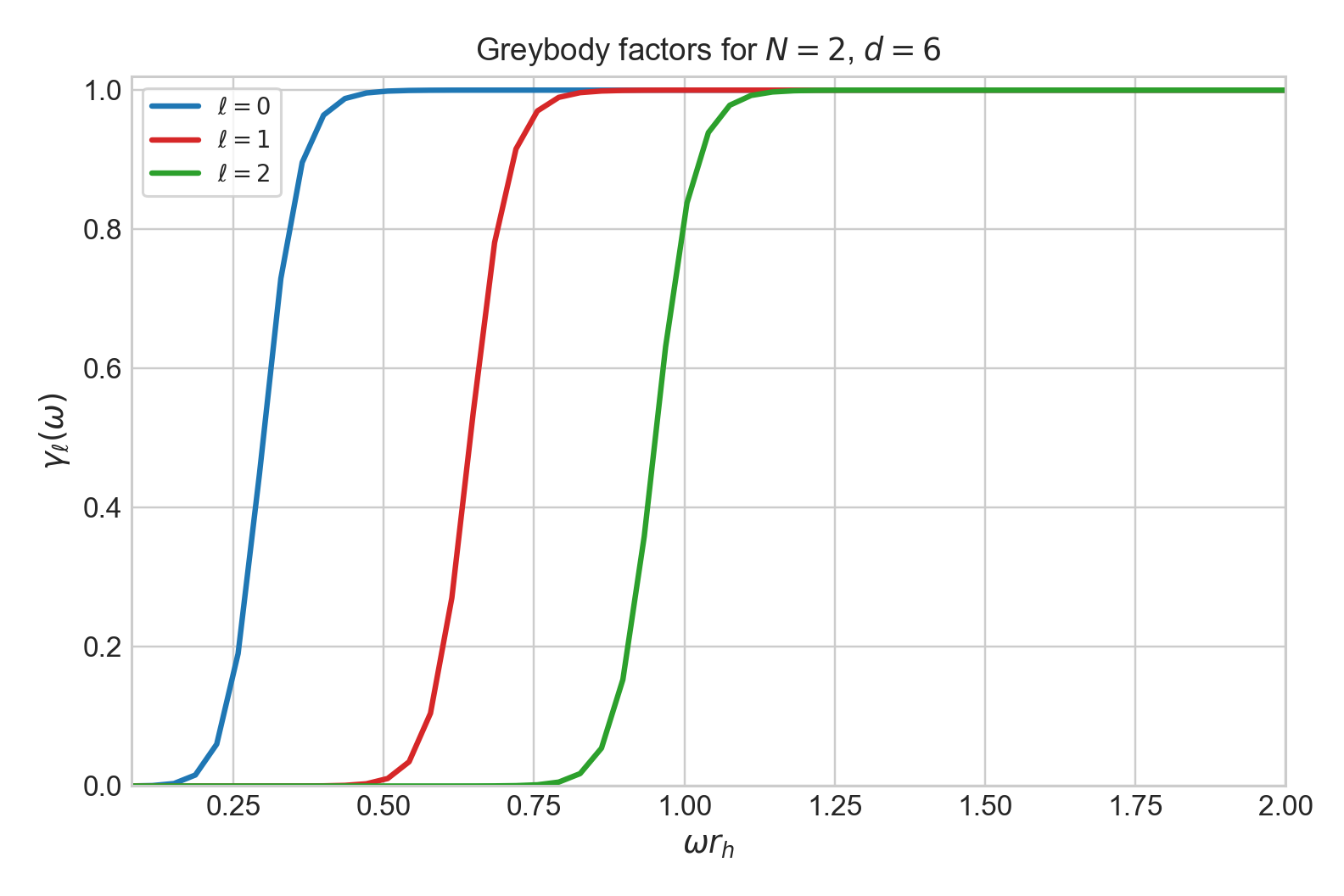}
\caption{Greybody factors of the minimally coupled test scalar field for the pure Lovelock black hole with $N=2$ and $d=6$. The transmission probability is shifted to higher frequencies as the angular momentum number increases.}
\label{fig:n2_d6_gbf}
\end{figure}

The corresponding greybody factors are plotted in Fig.~\ref{fig:n2_d6_gbf}. The curves show a clean barrier-penetration pattern: the $s$-wave turns on first, while the $\ell=1$ and $\ell=2$ modes require appreciably larger frequencies before reaching significant transmission. Quantitatively, the condition $\gamma_{\ell}\simeq 1/2$ occurs at about $\omega r_h\simeq 0.30$ for $\ell=0$, $\omega r_h\simeq 0.64$ for $\ell=1$, and $\omega r_h\simeq 0.95$ for $\ell=2$. Hence the low-frequency Hawking flux is expected to be dominated overwhelmingly by the $s$-wave, while higher multipoles only become important once the frequency is comparable to the top of the corresponding barrier.

\subsection{Representative Scalar Absorption Cross Sections}

\begin{figure}[t]
\centering
\includegraphics[width=0.82\textwidth]{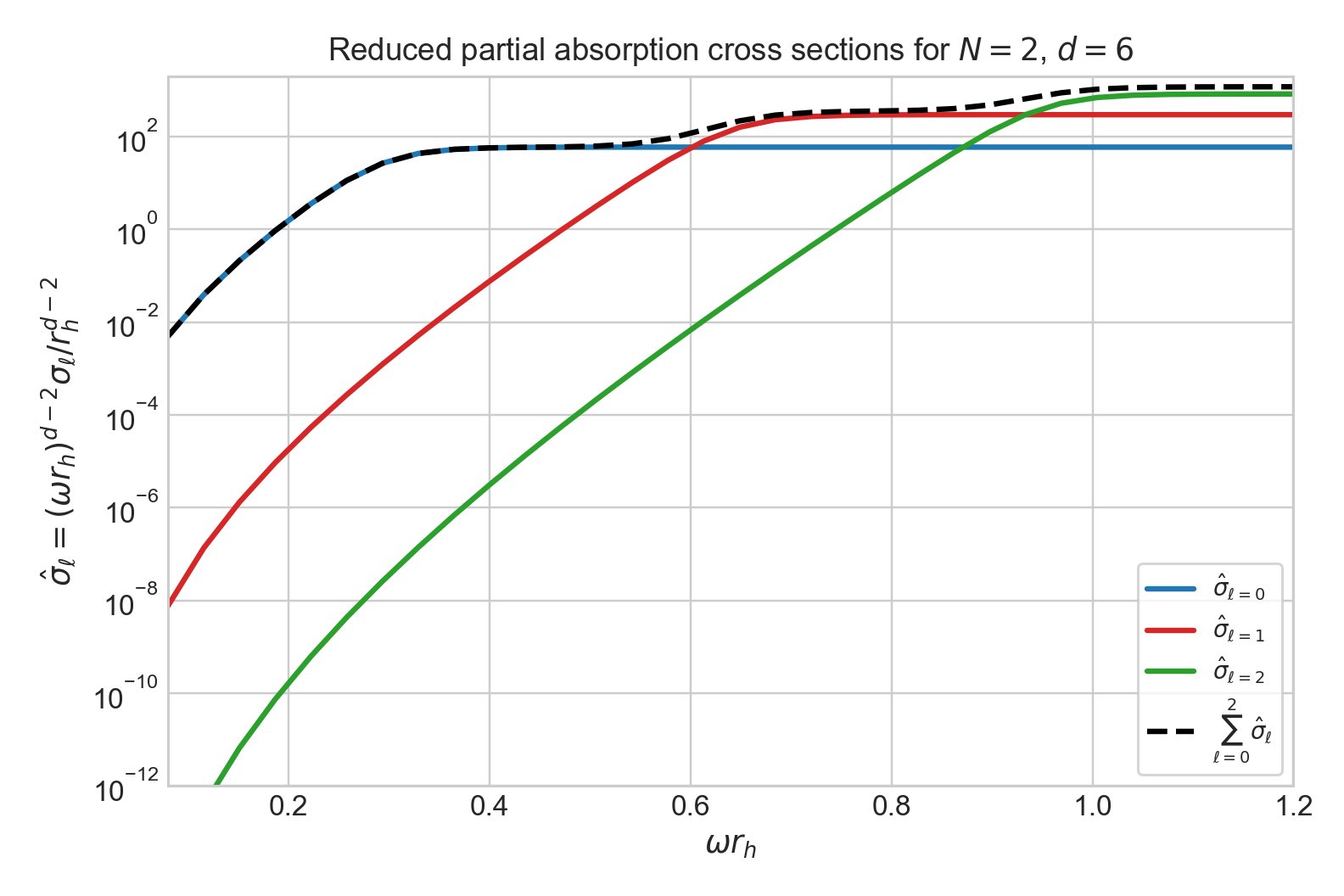}
\caption{Reduced partial scalar absorption cross sections for the pure Lovelock black hole with $N=2$ and $d=6$. The dashed curve shows the representative partial sum $\sum_{\ell=0}^{2}\hat{\sigma}_\ell$. In the low-frequency regime absorption is dominated by the $s$-wave, while higher multipoles become important only after the corresponding barriers are substantially penetrated.}
\label{fig:n2_d6_absorption}
\end{figure}

Having identified how the scalar greybody factors turn on, we next translate the same information into absorption observables. The partial absorption cross sections follow directly from Eqs.~\eqref{eq:partial_sigma_abs} and \eqref{eq:reduced_sigma_abs}. Since the first few partial waves dominate the low- and intermediate-frequency regime, it is sufficient for a representative discussion to display $\hat{\sigma}_{\ell}$ for $\ell=0,1,2$ in the benchmark case $N=2$, $d=6$. Figure~\ref{fig:n2_d6_absorption} shows these reduced partial cross sections together with their truncated sum $\sum_{\ell=0}^{2}\hat{\sigma}_\ell$.

Two features are worth emphasizing. First, the $s$-wave controls the absorption very close to threshold because the higher multipoles are exponentially suppressed by the barrier. Second, once the frequency increases, the larger harmonic multiplicities compensate for the higher barriers, so the $\ell=1$ contribution overtakes the $s$-wave around $\omega r_h\approx 0.60$, while the $\ell=2$ mode becomes comparable only closer to $\omega r_h\approx 0.9$. This is precisely the pattern expected from partial-wave scattering: barrier suppression governs the infrared regime, whereas degeneracy factors become increasingly important as the black hole becomes more transparent.

\subsection{Higher-Order Pure Lovelock Scalar Cases}

To assess how the pure Lovelock order affects the scalar greybody spectrum, we next compare the critical sequence $(N,d)=(2,6)$, $(3,8)$ and $(4,10)$. Restricting to the $s$-wave isolates the genuinely radial part of the scattering problem and makes the dependence on $N$ easiest to interpret.

\begin{figure}[t]
\centering
\includegraphics[width=\textwidth]{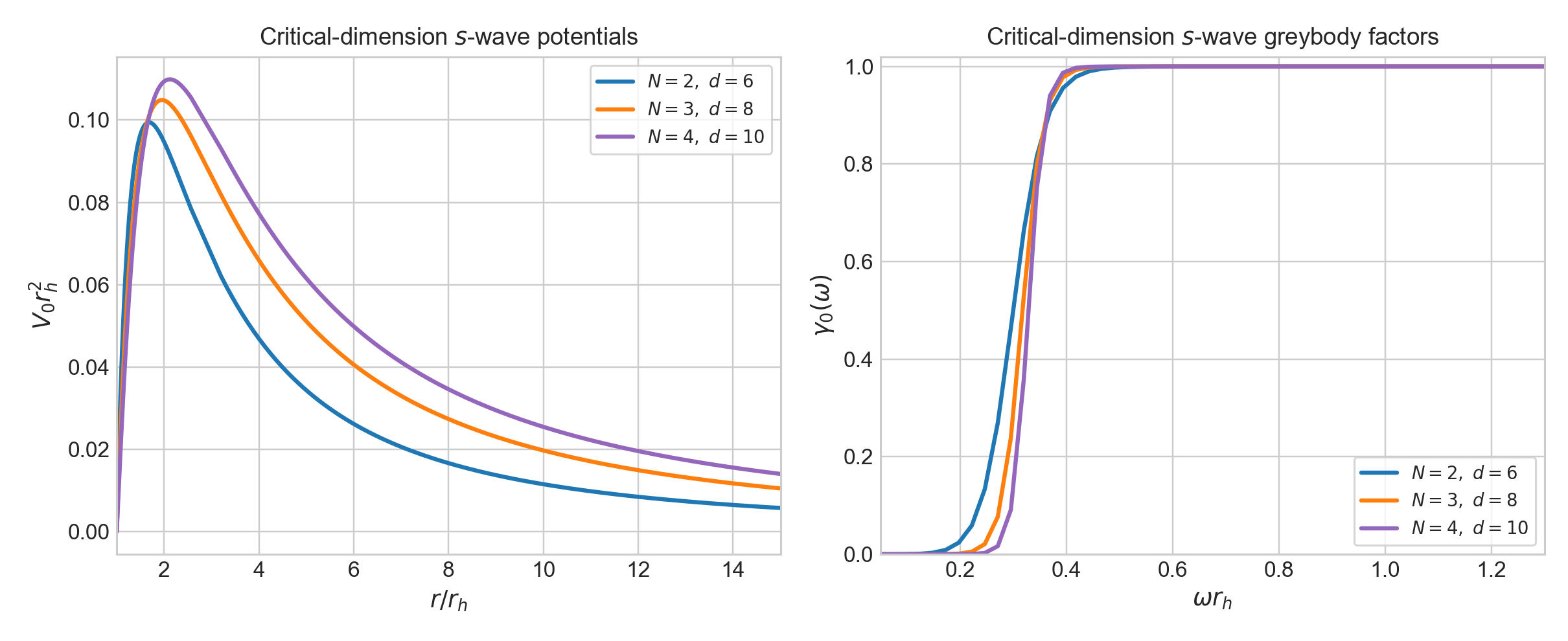}
\caption{Comparison of the critical pure Lovelock sequence $(N,d)=(2,6)$, $(3,8)$ and $(4,10)$ for the scalar $s$-wave. Left: effective potential. Right: greybody factor. The higher-order cases show a slightly higher and more extended barrier, together with a modest rightward shift of the transmission curve.}
\label{fig:critical_branch_compare}
\end{figure}

Figure~\ref{fig:critical_branch_compare} summarizes the result. The left panel shows that increasing $N$ raises the peak of the $s$-wave potential only mildly, from $V_{0,\max}r_h^2\simeq 0.0994$ for $N=2$ to $0.1049$ for $N=3$ and $0.1099$ for $N=4$, while the peak position moves outward from $r/r_h\simeq 1.68$ to $1.95$ and $2.13$, respectively. The right panel shows the associated greybody factors. The turn-on of the transmission curve is displaced to slightly higher frequencies as the Lovelock order increases, but the effect remains moderate on the critical branch: the representative value $\gamma_0\simeq 1/2$ occurs at $\omega r_h\simeq 0.300$ for $N=2$, $0.317$ for $N=3$, and $0.328$ for $N=4$.

The higher-$N$ comparison reveals that, once the theory is restricted to its dynamically preferred critical dimensions, increasing the Lovelock order does not qualitatively alter the scalar scattering problem. Instead, it produces a controlled quantitative suppression of the low-frequency transmission. The barrier grows, the onset of transmission is delayed, but the approach to unit transmission remains rapid once $\omega r_h$ exceeds the peak scale.

\subsection{Scalar Emissivity and Comparison with Tangherlini}

To place the pure Lovelock results in a more familiar higher-dimensional context, we compare them with the Schwarzschild--Tangherlini metric in the same dimension,
\begin{equation}
\label{eq:tangherlini_lapse}
f_{\mathrm{T}}(r)=1-\left(\frac{r_h}{r}\right)^{d-3}.
\end{equation}
The subscript $\mathrm{T}$ labels the Tangherlini benchmark, i.e. the higher-dimensional Schwarzschild solution of Einstein gravity.
In the benchmark dimension $d=6$, the pure Lovelock scalar barrier is much lower than the Tangherlini one: the peak values are $V_{0,\max}r_h^2\simeq 0.0994$ and $1.269$, respectively, and the corresponding half-transmission frequencies are $\omega r_h\simeq 0.300$ and $0.957$. Thus, at fixed horizon radius, the pure Lovelock black hole is geometrically more transparent even though its lapse function decays more slowly at infinity.

\begin{figure}[t]
\centering
\includegraphics[width=\textwidth]{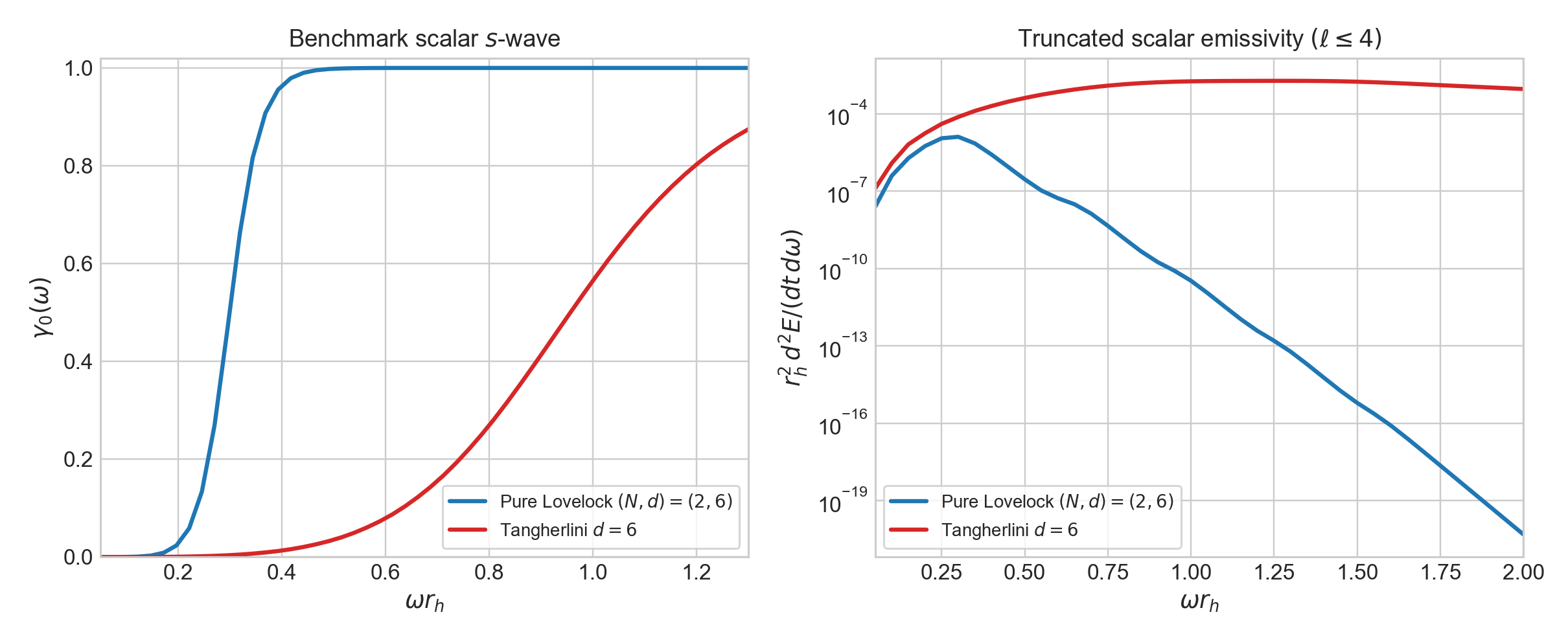}
\caption{Benchmark comparison of pure Lovelock and Tangherlini scalar scattering in $d=6$. Left: scalar $s$-wave greybody factor. Right: truncated scalar Hawking emissivity with $\ell\leq 4$. Pure Lovelock is more transparent at fixed $r_h$, but the much lower Hawking temperature suppresses the emitted spectrum.}
\label{fig:scalar_pure_vs_tangherlini_compare}
\end{figure}

This greater transparency does not translate into stronger Hawking emission because the pure Lovelock black hole is much colder. On the critical branch one has $T_H^{\mathrm{PL}}=1/(4\pi N r_h)$, whereas for Tangherlini $T_H^{\mathrm{T}}=(d-3)/(4\pi r_h)$. In the benchmark $d=6$ case this means $T_H^{\mathrm{PL}}=1/(8\pi)$ but $T_H^{\mathrm{T}}=3/(4\pi)$. Figure~\ref{fig:scalar_pure_vs_tangherlini_compare} makes the competition between geometry and thermodynamics explicit. The left panel shows the earlier turn-on of the pure Lovelock $s$-wave, while the right panel shows that the truncated scalar emissivity $\sum_{\ell=0}^{4} d^2E_\ell/(dt\,d\omega)$ is nevertheless much smaller. In the numerical spectra displayed here, the pure Lovelock scalar emissivity is suppressed by about three orders of magnitude in integrated strength relative to the Tangherlini case.

With the scalar benchmark established, we now turn to the Maxwell field to see whether the additional higher-dimensional polarizations merely decorate the scalar picture or sharpen it into a more distinctive pure Lovelock signature.

\subsection{Electromagnetic Benchmark and Critical Sequence}

\begin{figure}[t]
\centering
\includegraphics[width=\textwidth]{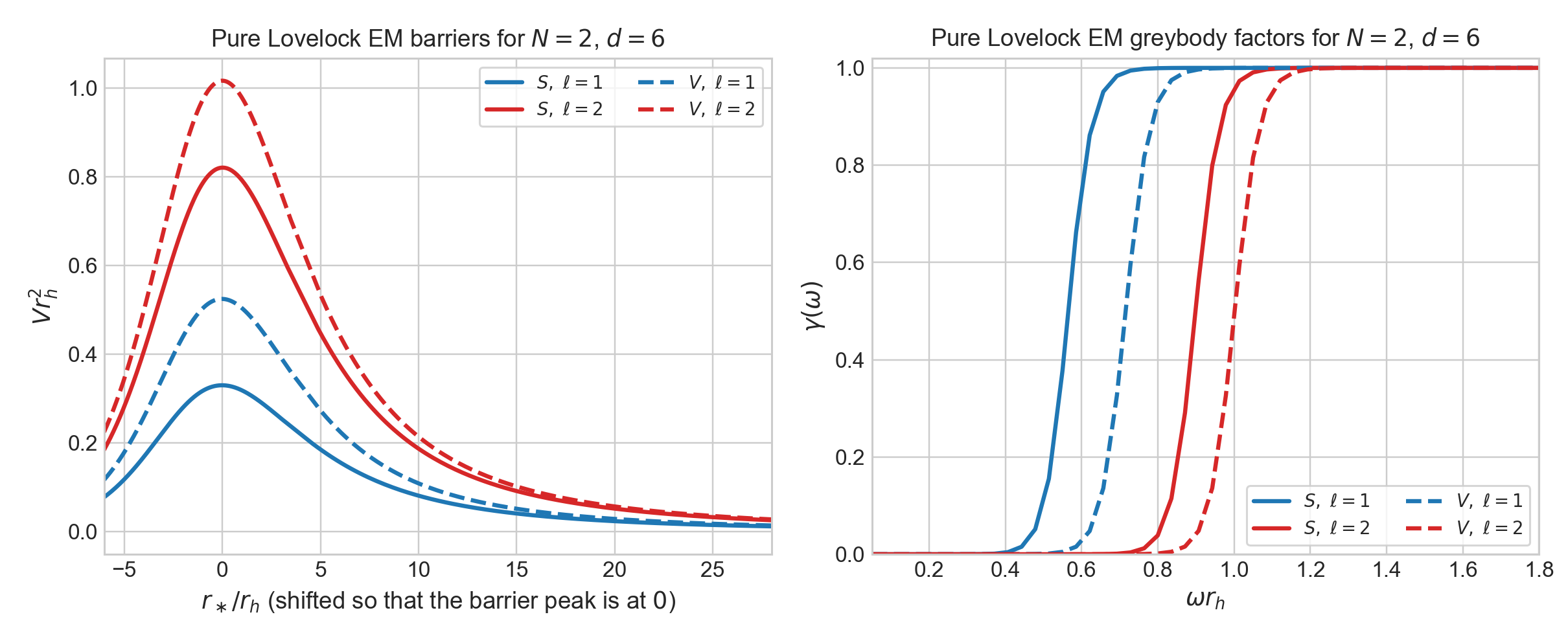}
\caption{Benchmark electromagnetic scattering for the pure Lovelock black hole with $N=2$ and $d=6$. Left: scalar-type ($S$) and vector-type ($V$) barriers for $\ell=1,2$. Right: corresponding greybody factors. At fixed multipole, the scalar-type channel is systematically more transparent than the vector-type one.}
\label{fig:em_n2_d6_overview}
\end{figure}

The higher-dimensional Maxwell field provides a qualitatively richer problem because it contains two physical sectors. Figure~\ref{fig:em_n2_d6_overview} shows the scalar-type and vector-type benchmark barriers for $(N,d)=(2,6)$ together with the corresponding greybody factors for $\ell=1,2$. At fixed $\ell$, the scalar-type potential is always lower than the vector-type one, so the scalar-type channel turns on first. For the lowest physical multipole $\ell=1$, the representative value $\gamma\simeq 1/2$ occurs at $\omega r_h\simeq 0.565$ for the scalar-type mode and $0.716$ for the vector-type mode; for $\ell=2$ the corresponding values are $0.899$ and $1.002$.

These results show that the higher-dimensional photon is not simply a minor variation on the scalar case. The electromagnetic lowest mode already begins at $\ell=1$, and the sector splitting creates an internal hierarchy: the scalar-type channel controls the low-frequency electromagnetic transmission, while the vector-type channel remains noticeably more suppressed. This splitting disappears only in four dimensions, so it is a genuinely higher-dimensional feature.

\begin{figure}[t]
\centering
\includegraphics[width=\textwidth]{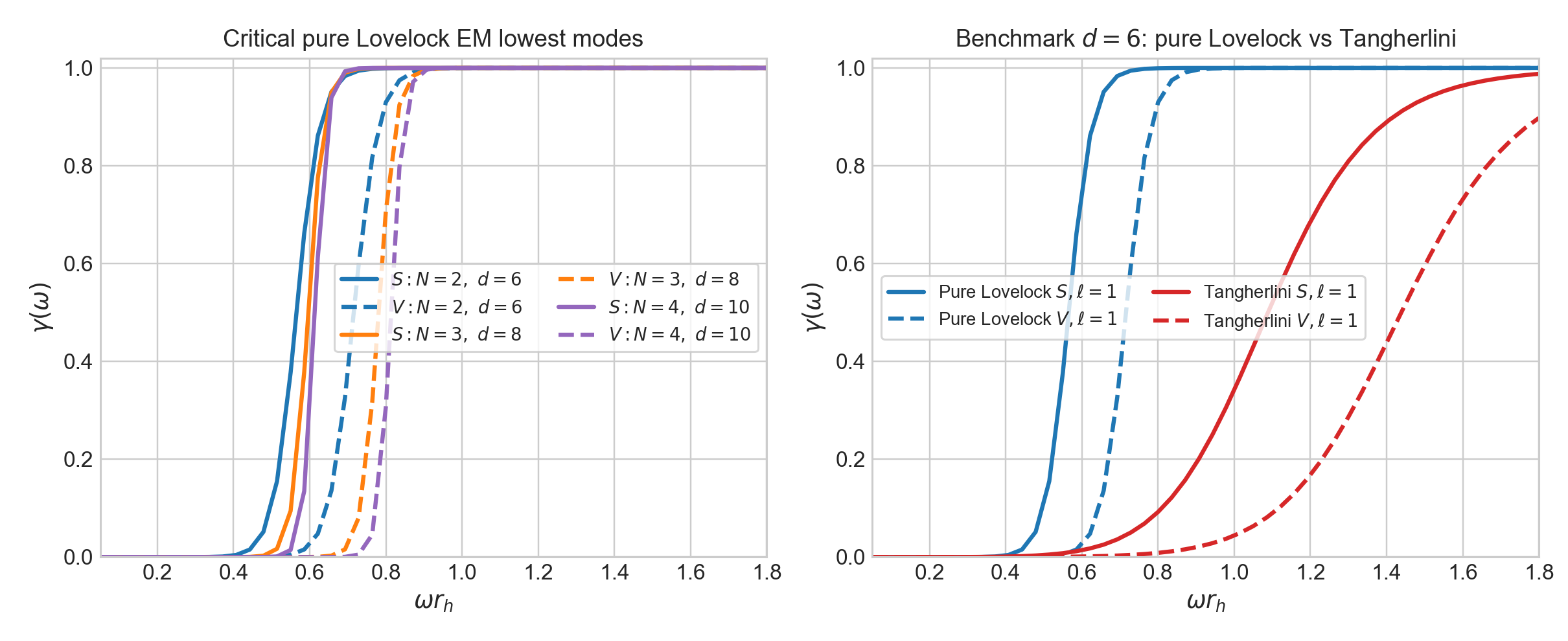}
\caption{Electromagnetic greybody comparisons. Left: lowest scalar-type and vector-type electromagnetic modes along the critical pure Lovelock sequence $(N,d)=(2,6)$, $(3,8)$ and $(4,10)$. Right: benchmark $d=6$ comparison between pure Lovelock and Tangherlini black holes. The pure Lovelock photon channels turn on substantially earlier than the Tangherlini ones.}
\label{fig:em_comparison_overview}
\end{figure}

Figure~\ref{fig:em_comparison_overview} extends the comparison in two directions. The left panel tracks the lowest electromagnetic modes along the critical pure Lovelock sequence $(N,d)=(2,6)$, $(3,8)$ and $(4,10)$. The scalar-type half-transmission point shifts only mildly, from $\omega r_h\simeq 0.565$ to $0.597$ and $0.613$, while the vector-type one moves from $0.716$ to $0.781$ and $0.814$. Thus, even for the electromagnetic field, the critical pure Lovelock sequence deforms the spectrum only smoothly as $N$ increases.

The right panel of Fig.~\ref{fig:em_comparison_overview} compares the $d=6$ pure Lovelock benchmark with the Tangherlini black hole. The contrast is much stronger than in the pure Lovelock sequence itself: the Tangherlini half-transmission points occur at $\omega r_h\simeq 1.089$ for the scalar-type $\ell=1$ mode and $1.438$ for the vector-type one. In other words, the Tangherlini electromagnetic barriers are substantially more opaque at fixed $d$ and $r_h$, in qualitative agreement with the existing higher-dimensional Schwarzschild literature \cite{JungPark2007}.

\subsection{Electromagnetic Emissivity and a Distinctive Pure Lovelock Signature}

\begin{figure}[t]
\centering
\includegraphics[width=\textwidth]{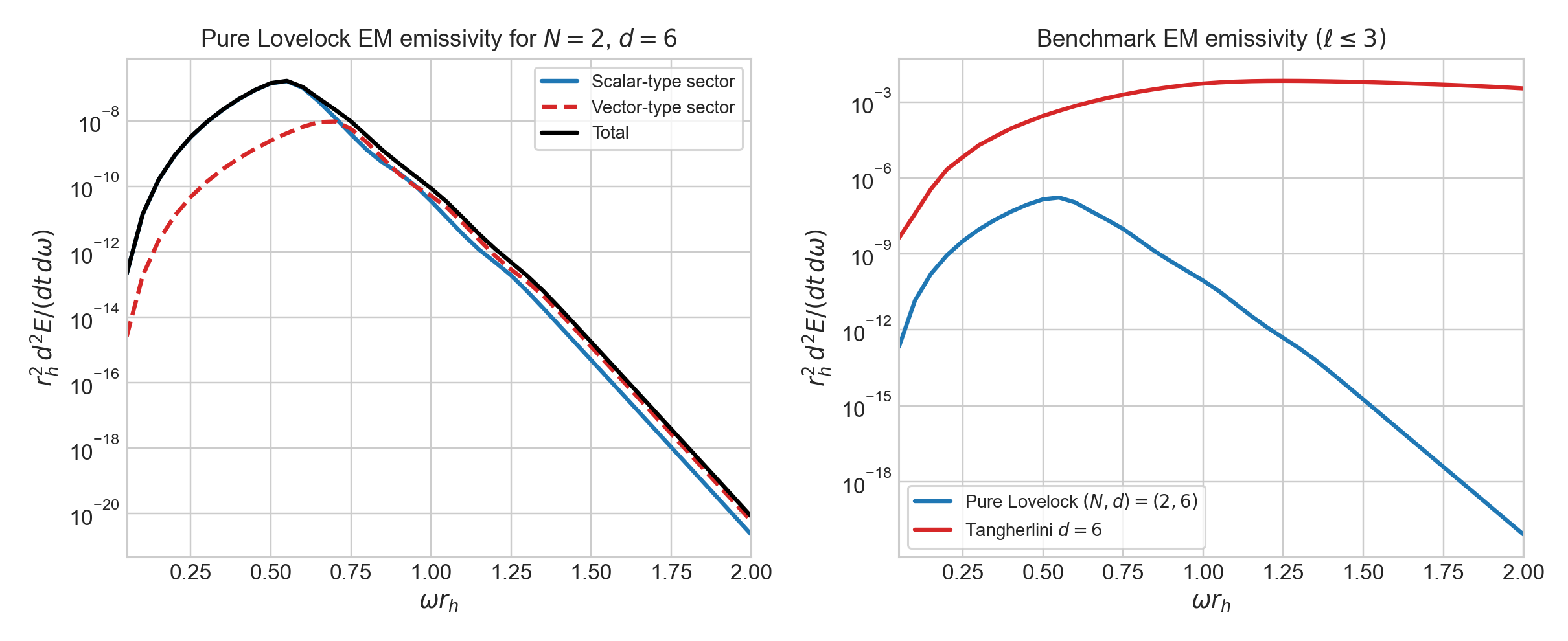}
\caption{Electromagnetic Hawking emissivity. Left: scalar-type, vector-type and total benchmark pure Lovelock spectra for $(N,d)=(2,6)$, truncated at $\ell\leq 3$. Right: total benchmark emissivity for the pure Lovelock and Tangherlini black holes in $d=6$. The pure Lovelock spectrum is much smaller despite the earlier turn-on of the greybody factors.}
\label{fig:em_emission_compare}
\end{figure}

The electromagnetic emissivity reveals the most distinctive pure Lovelock behavior. Figure~\ref{fig:em_emission_compare} (left) shows the benchmark pure Lovelock spectrum resolved into scalar-type and vector-type sectors. The scalar-type channel dominates the low-lying emission throughout the frequency window shown, while the vector-type contribution is subleading by almost an order of magnitude even near its own peak. For the first few benchmark multipoles included in the plot, the integrated vector-to-scalar emissivity ratio is only about $6.9\times 10^{-2}$.

This scalar-type dominance becomes even more pronounced along the critical pure Lovelock sequence. For the first two partial waves, the vector-to-scalar emissivity ratio drops from $6.9\times 10^{-2}$ at $(N,d)=(2,6)$ to $4.5\times 10^{-3}$ at $(3,8)$ and $3.0\times 10^{-4}$ at $(4,10)$. The reason is simple: the Hawking temperature decreases like $1/N$, while the vector-type threshold shifts steadily to the right. As a result, the higher-$N$ pure Lovelock photon output is increasingly forced into the scalar-type lowest modes.

The right panel of Fig.~\ref{fig:em_emission_compare} compares the benchmark total electromagnetic emissivity with the Tangherlini one. Despite the larger pure Lovelock greybody factors, the thermal suppression dominates overwhelmingly: the benchmark pure Lovelock spectrum peaks at $\omega r_h\simeq 0.55$ with height $\simeq 1.7\times 10^{-7}$, whereas the Tangherlini spectrum peaks near $\omega r_h\simeq 1.25$ with height $\simeq 6.9\times 10^{-3}$. Thus the pure Lovelock electromagnetic peak is lower by roughly $4\times 10^{4}$. This competition between enhanced transparency and much lower temperature is the clearest scattering signature of the pure Lovelock geometry.

Because the difference is so large already at the spectral level, it is natural to compress the comparison into integrated power estimates, which is the purpose of the next section.

\subsection{WKB Cross-Check of the Pure Lovelock Greybody Factors}

To check the direct integration against semi-analytic approximations, we evaluated the barrier-top WKB transmission formula implemented in the online WKB calculator used in the recent higher-order WKB greybody code of Konoplya, Matyjasek and Zhidenko \cite{KonoplyaMatyjasekZhidenko2026}, comparing the first-, third- and sixth-order continuations, with the third-order term built on the Iyer--Will framework \cite{IyerWill1987}. For each pure Lovelock mode we located the maximum of the effective potential in the tortoise coordinate, computed the derivatives $V_0^{(n)}=d^nV/dr_\ast^n\vert_{r_{\ast,0}}$ up to $n=12$, solved the continued WKB equation for the barrier parameter $\nu(\omega)$, and reconstructed the transmission probability as
\begin{equation}
\gamma^{\mathrm{WKB}}_{\ell}(\omega)=\frac{1}{1+e^{2\pi \nu(\omega)}}.
\end{equation}
At first order the half-transmission point is simply $\omega_{1/2}^{(1)}r_h=\sqrt{V_{0,\max}r_h^2}$, so the comparison also shows how much the higher-order WKB correction shifts the onset of transmission away from the naive barrier-top estimate.

This WKB cross-check also fits naturally into a growing body of recent work, where related WKB-based greybody analyses were carried out for Euler--Heisenberg black holes, Proca--Gauss--Bonnet black holes, Bonanno--Reuter regular black holes, four-dimensional quasi-topological gravity, and in the higher-order WKB code study itself \cite{Malik2025EEH,Lutfuoglu2025PGB,BolokhovSkvortsova2025BR,Dubinsky2026QTG,KonoplyaMatyjasekZhidenko2026}. Beyond these examples, WKB-based greybody calculations have likewise been used in dRGT massive gravity and in asymptotically safe or quantum-corrected black-hole backgrounds, while the same transmission/reflection machinery has also appeared in holographic conductivity calculations through the reflection coefficient of an effective Schr\"odinger barrier \cite{BoonsermEtAl2023dRGT,LutfuogluEtAl2025PT,Konoplya:2009hv}.

\begin{table}[t]
\centering
\small
\resizebox{\textwidth}{!}{%
\begin{tabular}{lccccccccc}
\hline
Mode & $\omega_{1/2}^{\mathrm{num}}r_h$ & $\omega_{1/2}^{(1)}r_h$ & $\omega_{1/2}^{(3)}r_h$ & $\omega_{1/2}^{(6)}r_h$ & $\delta_1$ [\%] & $\delta_3$ [\%] & $\delta_6$ [\%] & $\gamma^{(3)}(\omega_{1/2}^{\mathrm{num}})$ & $\gamma^{(6)}(\omega_{1/2}^{\mathrm{num}})$ \\
\hline
Scalar $(N,\ell)=(2,0)$ & 0.300 & 0.315 & 0.299 & 0.299 & +5.12 & -0.42 & -0.23 & 0.511 & 0.506 \\
Scalar $(N,\ell)=(2,1)$ & 0.640 & 0.653 & 0.644 & 0.644 & +2.01 & +0.67 & +0.68 & 0.466 & 0.466 \\
Scalar $(N,\ell)=(2,2)$ & 0.950 & 0.958 & 0.952 & 0.952 & +0.83 & +0.21 & +0.21 & 0.484 & 0.484 \\
Scalar $(N,\ell)=(3,0)$ & 0.317 & 0.324 & 0.317 & 0.315 & +2.15 & +0.11 & -0.62 & 0.495 & 0.525 \\
Scalar $(N,\ell)=(4,0)$ & 0.328 & 0.332 & 0.328 & 0.328 & +1.07 & +0.04 & +0.11 & 0.498 & 0.494 \\
EM scalar-type $(N,\ell)=(2,1)$ & 0.565 & 0.574 & 0.565 & 0.565 & +1.64 & +0.02 & +0.03 & 0.499 & 0.498 \\
EM vector-type $(N,\ell)=(2,1)$ & 0.716 & 0.724 & 0.716 & 0.716 & +1.13 & +0.00 & +0.01 & 0.500 & 0.499 \\
EM scalar-type $(N,\ell)=(3,1)$ & 0.597 & 0.600 & 0.596 & 0.596 & +0.48 & -0.18 & -0.23 & 0.513 & 0.516 \\
EM vector-type $(N,\ell)=(3,1)$ & 0.781 & 0.784 & 0.781 & 0.781 & +0.43 & -0.00 & -0.00 & 0.500 & 0.500 \\
EM scalar-type $(N,\ell)=(4,1)$ & 0.613 & 0.617 & 0.614 & 0.614 & +0.57 & +0.22 & +0.20 & 0.478 & 0.480 \\
EM vector-type $(N,\ell)=(4,1)$ & 0.814 & 0.815 & 0.813 & 0.813 & +0.14 & -0.08 & -0.08 & 0.510 & 0.510 \\
\hline
\end{tabular}%
}
\caption{Comparison between the manuscript's direct-integration half-transmission points and the WKB estimates for representative pure Lovelock modes. Here $\delta_n=100\,\bigl(\omega_{1/2}^{(n)}-\omega_{1/2}^{\mathrm{num}}\bigr)/\omega_{1/2}^{\mathrm{num}}$, and the last two columns show how close the third- and sixth-order WKB approximations come to the expected value $\gamma\simeq 1/2$ at the quoted numerical transition frequency.}
\label{tab:wkb_halfpoints}
\end{table}

Table~\ref{tab:wkb_halfpoints} shows that the first-order WKB estimate systematically overshoots the turn-on frequency, by $5.1\%$ for the benchmark scalar $s$-wave and by $0.1\%$--$2.2\%$ across the rest of the sample. Both the third- and sixth-order corrections keep the discrepancy below $0.7\%$ for every pure Lovelock mode listed, and below $0.25\%$ for nine of the eleven entries. The sixth-order approximation is therefore comparably accurate at the quoted half-transmission points, although not uniformly closer there: for instance, the scalar $(N,\ell)=(3,0)$ mode shifts from $+0.11\%$ at third order to $-0.62\%$ at sixth order. At the quoted numerical half-transmission frequencies the third-order WKB values lie between $0.466$ and $0.513$, whereas the sixth-order ones lie between $0.466$ and $0.525$.

\begin{table}[t]
\centering
\small
\resizebox{\textwidth}{!}{%
\begin{tabular}{lcccccc}
\hline
Mode & $\omega r_h$ & $\gamma^{\mathrm{num}}$ & $\gamma^{(3)}_{\mathrm{WKB}}$ & $\gamma^{(6)}_{\mathrm{WKB}}$ & $\lvert\Delta\gamma\rvert_{(3)}$ & $\lvert\Delta\gamma\rvert_{(6)}$ \\
\hline
Scalar $N=2$, $\ell=0$ & 0.200 & 0.026735 & 0.025393 & 0.026912 & 0.001342 & 0.000177 \\
Scalar $N=2$, $\ell=0$ & 0.300 & 0.505726 & 0.510758 & 0.505860 & 0.005032 & 0.000134 \\
Scalar $N=2$, $\ell=0$ & 0.400 & 0.963920 & 0.963631 & 0.963986 & 0.000289 & 0.000066 \\
EM scalar-type $N=2$, $\ell=1$ & 0.450 & 0.019868 & 0.019716 & 0.019874 & 0.000152 & 0.000005 \\
EM scalar-type $N=2$, $\ell=1$ & 0.565 & 0.498414 & 0.499257 & 0.498388 & 0.000843 & 0.000026 \\
EM scalar-type $N=2$, $\ell=1$ & 0.700 & 0.986631 & 0.986569 & 0.986631 & 0.000061 & 0.000001 \\
EM vector-type $N=2$, $\ell=1$ & 0.600 & 0.024655 & 0.024601 & 0.024656 & 0.000054 & 0.000001 \\
EM vector-type $N=2$, $\ell=1$ & 0.716 & 0.499342 & 0.499903 & 0.499338 & 0.000560 & 0.000005 \\
EM vector-type $N=2$, $\ell=1$ & 0.850 & 0.983371 & 0.983348 & 0.983370 & 0.000023 & 0.000001 \\
\hline
\end{tabular}%
}
\caption{Benchmark $d=6$ pure Lovelock greybody factors from direct numerical integration and from the third- and sixth-order WKB approximations. The sample points are chosen to cover the onset, transition and near-transparent regimes of the lowest scalar and electromagnetic channels.}
\label{tab:wkb_samples}
\end{table}

Table~\ref{tab:wkb_samples} shows that the agreement is not limited to a single transition point. For the benchmark scalar and electromagnetic lowest modes, the maximum absolute difference drops from $5.0\times10^{-3}$, $8.4\times10^{-4}$ and $5.6\times10^{-4}$ at third order to $1.8\times10^{-4}$, $2.6\times10^{-5}$ and $4.6\times10^{-6}$ at sixth order for the scalar $s$-wave, scalar-type photon and vector-type photon, respectively. Over seven equally spaced points in the windows $\omega r_h\in[\omega_{1/2}-0.15,\omega_{1/2}+0.15]$, the mean absolute discrepancy drops from $1.57\times10^{-3}$ to $1.15\times10^{-4}$ for the scalar $s$-wave, from $2.50\times10^{-4}$ to $7.58\times10^{-6}$ for the scalar-type photon, and from $1.65\times10^{-4}$ to $1.90\times10^{-6}$ for the vector-type photon. Thus the sixth-order WKB treatment reproduces both the onset frequencies and the local shape of the benchmark pure Lovelock greybody curves substantially more accurately in the transition region most relevant for Hawking emission.

\section{Estimated Hawking Energy Emission Rates}

The spectral curves discussed above can now be condensed into explicit power estimates by integrating the Hawking energy-emission formulas over frequency. To keep the numerics transparent, we work with truncated sums, where $\ell_{\max}$ denotes the highest multipole retained in the partial-wave expansion. In the dimensionless units used throughout the paper, with $r_h=1$, we define
\begin{equation}
\label{eq:truncated_scalar_power}
\mathcal{P}^{\leq \ell_{\max}}_{\mathrm{sc}}=\int_0^{\infty}\left.\frac{d^2E_{\mathrm{sc}}}{dt\,d\omega}\right|_{\ell\leq \ell_{\max}}d\omega,
\end{equation}
for the scalar field, and
\begin{equation}
\label{eq:truncated_em_power}
\mathcal{P}^{\leq \ell_{\max}}_{\mathrm{EM}}=\int_0^{\infty}\left.\frac{d^2E_{\mathrm{EM}}}{dt\,d\omega}\right|_{\ell\leq \ell_{\max}}d\omega,
\end{equation}
for the electromagnetic field. In practice we use $\ell_{\max}=4$ for the scalar field and $\ell_{\max}=3$ for the electromagnetic field. These choices are sufficient for a stable first estimate because the spectra are strongly concentrated in the lowest partial waves: the Hawking temperature is low on the pure Lovelock critical branch, and the higher multipoles only turn on after appreciable barrier penetration. Accordingly, the integrated powers reported below should be read as truncated test-field benchmarks at fixed horizon radius rather than as complete equal-mass luminosities; the most robust outputs are the relative suppression factors and the reorganization of the electromagnetic sector hierarchy.

\begin{table}[t]
\centering
\small
\begin{tabular}{lcccc}
\hline
Background & $\mathcal{P}^{\leq 4}_{\mathrm{sc}}$ & $\mathcal{P}^{\leq 3}_{\mathrm{EM},S}$ & $\mathcal{P}^{\leq 3}_{\mathrm{EM},V}$ & $\mathcal{P}^{\leq 3}_{\mathrm{EM}}$ \\
\hline
Pure Lovelock $(2,6)$ & $2.16\times 10^{-6}$ & $3.16\times 10^{-8}$ & $2.20\times 10^{-9}$ & $3.38\times 10^{-8}$ \\
Pure Lovelock $(3,8)$ & $1.95\times 10^{-8}$ & $9.12\times 10^{-12}$ & $4.14\times 10^{-14}$ & $9.16\times 10^{-12}$ \\
Pure Lovelock $(4,10)$ & $1.72\times 10^{-10}$ & $2.16\times 10^{-15}$ & $6.40\times 10^{-19}$ & $2.16\times 10^{-15}$ \\
Tangherlini $d=6$ & $2.23\times 10^{-3}$ & $4.13\times 10^{-3}$ & $2.72\times 10^{-3}$ & $6.85\times 10^{-3}$ \\
\hline
\end{tabular}
\caption{Truncated Hawking power estimates obtained by integrating the spectra generated from Eqs.~\eqref{eq:scalar_emission} and \eqref{eq:em_emission}. The scalar estimates include modes up to $\ell=4$, while the electromagnetic estimates include modes up to $\ell=3$. All numbers are quoted in the dimensionless units used in the numerical analysis with $r_h=1$.}
\label{tab:power_estimates}
\end{table}

Table~\ref{tab:power_estimates} makes the thermal suppression of the pure Lovelock branch very explicit. In the benchmark $d=6$ case, the truncated scalar power is $\mathcal{P}^{\leq 4}_{\mathrm{sc}}\simeq 2.16\times10^{-6}$ for pure Lovelock, compared with $2.23\times10^{-3}$ for Tangherlini. Thus the benchmark pure Lovelock scalar power is smaller by a factor of about $9.7\times10^{-4}$ even though the corresponding greybody factors are larger at low and intermediate frequencies. The same competition is even more dramatic in the electromagnetic sector: the truncated pure Lovelock power is only $\mathcal{P}^{\leq 3}_{\mathrm{EM}}\simeq 3.38\times10^{-8}$, whereas the Tangherlini value is $6.85\times10^{-3}$, yielding a suppression factor of about $4.9\times10^{-6}$.

The sector-resolved electromagnetic entries show quantitatively how the pure Lovelock photon output is reorganized. For the benchmark pure Lovelock black hole, the ratio of vector-type to scalar-type emitted power is only $\mathcal{P}^{\leq 3}_{\mathrm{EM},V}/\mathcal{P}^{\leq 3}_{\mathrm{EM},S}\simeq 6.95\times10^{-2}$, while for Tangherlini the corresponding ratio is $\simeq 6.58\times10^{-1}$. In other words, the scalar-type channel dominates much more strongly in pure Lovelock than in the Einstein benchmark. This is the integrated version of the behavior already visible in the greybody factors and emissivity spectra.

Along the critical pure Lovelock sequence, the total power decreases very rapidly with Lovelock order. For the scalar field, the truncated power falls from $2.16\times10^{-6}$ at $(N,d)=(2,6)$ to $1.95\times10^{-8}$ at $(3,8)$ and $1.72\times10^{-10}$ at $(4,10)$, i.e. by roughly two orders of magnitude at each step. The electromagnetic power drops even faster, from $3.38\times10^{-8}$ to $9.16\times10^{-12}$ and then to $2.16\times10^{-15}$. This confirms that once the Hawking temperature and the higher-mode thresholds are folded together, the observable energy flux is suppressed far more strongly than one would infer from the greybody factors alone.

\section{Conclusion}

We have followed a single physical thread from effective potentials to greybody factors, frequency-resolved Hawking spectra and finally integrated power estimates for asymptotically flat pure Lovelock black holes. The main technical result is that both the minimally coupled scalar field and the higher-dimensional Maxwell field reduce to one-dimensional barrier problems on the critical branch $d=2N+2$, so the role of each quantity is transparent: $N$ controls the higher-curvature order, $d$ fixes the spacetime dimension, $r_h$ sets the length scale, $\ell$ labels the angular multipole, and $\omega$ labels the emitted frequency. For the scalar field, increasing $N$ produces only a mild rightward shift of the transmission curve. For the Maxwell field, the scalar-type channel is always more transparent than the vector-type one, so the critical sequence deforms the spectrum smoothly but with a clear internal hierarchy.

The comparison with Schwarzschild--Tangherlini black holes reveals the central message of the paper most cleanly. At fixed spacetime dimension and horizon radius, pure Lovelock geometries are more transparent, so both scalar and electromagnetic greybody factors turn on earlier. But the same black holes are much colder, which means that thermodynamics dominates over transmission once Hawking radiation is considered. This is why the frequency-resolved spectra and the integrated powers are both strongly suppressed relative to Tangherlini, despite the lower barriers. In the electromagnetic sector the competition leaves an especially sharp imprint: as $N$ grows, the emitted power is pushed more and more strongly into the scalar-type channel. Taken together, the barrier data, emissivity curves and power estimates show that ``more transparent but colder'' is the unifying Hawking-radiation signature of pure Lovelock black holes. This conclusion should be understood within the deliberately controlled comparison adopted here: fixed background, fixed horizon radius and truncated multipole sums. Within that scope, the neglected backreaction and residual high-$\ell$ tails are not expected to overturn the qualitative ordering reported in the paper.

\end{document}